\documentclass[twocolumn,prb,superscriptaddress,noshowpacs,epsf]{revtex4}
\usepackage{graphicx}
\usepackage{epsf}
\usepackage{amsmath}

\begin{document}

\title{Cooling a mechanical resonator via coupling to a tunable double quantum dot}
\date{\today}
\author{Shi-Hua Ouyang}
\affiliation{Department of Physics and Surface Physics Laboratory
(National Key Laboratory), Fudan University, Shanghai 200433, China}
\affiliation{Advanced Study Institute, The Institute of Physical and
Chemical Research (RIKEN), Wako-shi 351-0198, Japan}
\author{J. Q. You}
\affiliation{Department of Physics and Surface Physics Laboratory
(National Key Laboratory), Fudan University, Shanghai 200433, China}
\affiliation{Advanced Study Institute, The Institute of Physical and
Chemical Research (RIKEN), Wako-shi 351-0198, Japan}
\author{Franco Nori}
\affiliation{Advanced Study Institute, The Institute of Physical and
Chemical Research (RIKEN), Wako-shi 351-0198, Japan}
\affiliation{Center for Theoretical Physics, Physics Department,
Center for the Study of Complex Systems, University of Michigan, Ann
Arbor, MI 48109-1040, USA}

\begin{abstract}
We study the cooling of a mechanical resonator (MR) that is
capacitively coupled to a double quantum dot (DQD). The MR is cooled
by the dynamical backaction induced by the capacitive coupling
between the DQD and the MR. The DQD is excited by a microwave field
and afterwards a tunneling event results in the decay of the excited
state of the DQD. An important advantage of this system is that both
the energy level splitting and the decay rate of the DQD can be well
tuned by varying the gate voltage. We find that the steady average
occupancy, below unity, of the MR can be achieved by changing both
the decay rate of the excited state and the detuning between the
transition frequency of the DQD and the microwave frequency, in
analogy to the laser sideband cooling of an atom or trapped ion in
atomic physics. Our results show that the cooling of the MR to the
ground state is experimentally implementable.
\end{abstract}

 \maketitle



\section{Introduction}
Mechanical resonators (MRs) are currently attracting considerable
interest because of their potential applications to high-precision
displacement detection,\cite{LaHaye04} mass detection,\cite{Buks06}
and quantum measurement.\cite{Braginsky92}  Recent technical
advances allow the fabrication of a MR with both a high quality
factor ($Q$-factor) and a sufficient high frequency, approaching
1~GHz.\cite{ANC96,Huang03,Gaidarzhy05} Such a MR provides a good
platform for exploring various quantum phenomena and for observing
the quantum-to-classical transition in macroscopic
objects.\cite{Schwab05,Wei06} Moreover, quantized MRs could be
useful in quantum information science. Indeed, the quantized motion
of buckling nanoscale bars has been proposed for implementing
qubits.\cite{Savel04,Savel07,SavelNJP} The generation of entangled
states,\cite{Armour02,Tian04} squeezed states,\cite{Xue07Squeezed}
and quantum nondemolition measurements\cite{Irish03} using MRs have
also been studied. However, to prepare an ideal ground state, the
basic requirement is to be able to cool the MR to a state with a
mean phonon number $\langle n\rangle\ll1$.

Numerous experiments cooling a single MR have been reported recently
(see, e.g.,
\onlinecite{Metzger04,Gigan06,Arcizet06,Kleckner06,Schliesser06,Poggio07,Lehnert08}).
In these experiments, the MR and a fixed micromirror form an optical
cavity and the MR is cooled by either radiation-pressure-induced
backaction or bolometric backaction. Experimental results show that
a single MR can be cooled down from room temperature to an effective
temperature of the order of 0.1~K (Refs.~\onlinecite{Gigan06} and
\onlinecite{Kleckner06}) or
10~K~(Refs.~\onlinecite{Metzger04,Arcizet06,Schliesser06}).
However, for a MR with frequency $\sim20$~MHz, a temperature lower
than $1$~mK is required in order to drive the MR to the quantum
regime. Thus, more effective cooling methods are needed, in addition
to increasing the oscillation frequency of the MR. Moreover, besides
the classical and semiclassical analyses of cooling a single MR via
dynamical backaction,\cite{Kippenberg05, Schliesser06,Braginsky02}
some quantum theories have also been
developed.\cite{Wilson07,Marquardt07,Genes08,Grajcar08} For example,
in Refs.~\onlinecite{Wilson07} and \onlinecite{Marquardt07}, it is
predicted that a MR can be cooled down to its ground state when the
frequency $\omega_m$ of the MR is either comparable to or larger
than the optical cavity's resonance linewidth $\Gamma$. In laser
cooling, this corresponds to the sideband cooling of a bound atom or
a trapped ion, where the lowest occupancy attainable is given by
$\langle n\rangle\approx \Gamma^2/16\omega_m^2\ll1$, indicating that
the MR can be most of the time in its ground state. Recently, the
resolved-sideband cooling of a MR has been realized by coupling the
MR to an optical resonant system.\cite{ASchliesser08}

Besides optomechanical cooling, an alternative way would be to cool
the MR by coupling it, via an electronic coupling, to an electronic
system. This provides the advantage of fabricating and integrating
the electronic device into a cryogenic system. In principle, the
electronic cooling of a MR can be achieved by several means,
including coupling the MR to: (1) an optical quantum
dot,\cite{Wilson04} (2) a Josephson-juction superconducting quantum
device\cite{Martin04,Zhang05,Valenzuela06,You08,Hauss08} (which
behaves like a superconducting artificial atom\cite{You05}), and (3)
a one-dimensional transmission line.\cite{Xue07} Moreover, the
experimental cooling of a MR, via coupling it to a superconductor
single-electron transistor\cite{Naik06} or to an {\it LC}
circuit,\cite{Brown07} has also been reported.

\subsection{Cooling a mechanical resonator coupled to a double quantum dot}

\begin{figure}[tbp]
\includegraphics[width=3.0in,
bbllx=157,bblly=252,bburx=461,bbury=650]{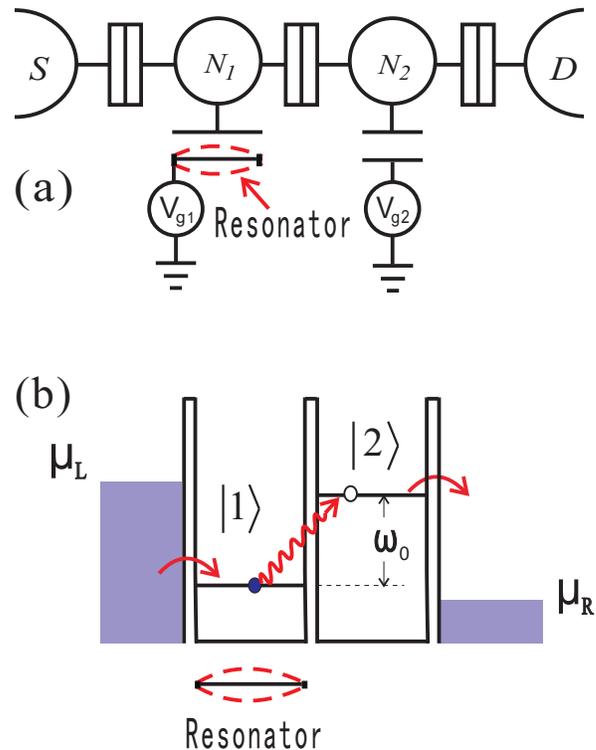} \caption{(Color
online)~(a) Schematic diagram of a DQD connected to an electron
source $S$ and drain $D$ via tunneling barriers. An oscillating
plate (the MR) and another static plate form a capacitor on the left
dot, which provides a capacitive coupling between the oscillating MR
and the DQD. The energy level of each dot is tunable by varying the
gate voltage $V_{g}$ applied to the dot through the capacitor. (b)
Transport process of an electron through a DQD: First, an electron
tunnels from the source to the left dot, and then a microwave field
drives it to the right dot. Finally, it tunnels to the drain on the
right side. \label{fig1}}
\end{figure}

In this work we propose an approach to cool a MR by coupling it to
an electronic system: a double quantum dot (DQD). Indeed, the whole
system consists of a DQD and a MR. The MR, together with another
static plate, forms a gate capacitor adjacent to the left dot (see
Fig.~\ref{fig1}). The oscillation of the MR will modulate the
effective capacitance of this capacitor. In this way, the MR can be
strongly coupled to the DQD.

\begin{figure}[tbp]
\includegraphics[width=3.20in,
bbllx=113,bblly=344,bburx=515,bbury=604]{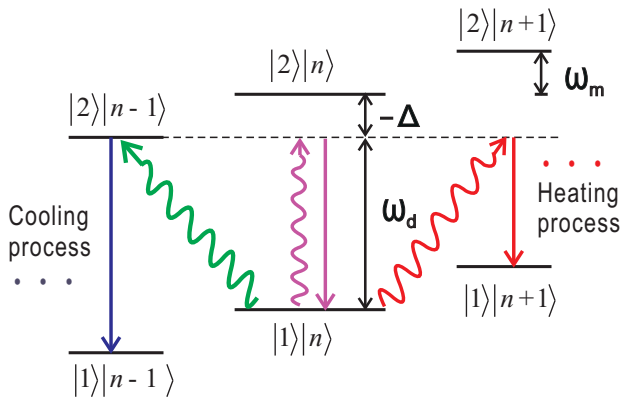} \caption{(Color
online)~Schematic diagram of the cooling (heating) process in the
coupled MR-DQD system. When the DQD is excited by a red-detuned
microwave field, i.e., $\Delta=\omega_d-\omega_0<0$, the anti-Stokes
process ($|1\rangle|n\rangle\rightarrow|2\rangle|n-1\rangle$) is
resonantly enhanced. A subsequent decay from the excited state
$|2\rangle$ to the ground state $|1\rangle$ reduces the energy of
the MR by one quanta. This cools the MR because the emitted (blue)
photon has more energy than the absorbed (green) photon. Due to the
off-resonance, the Stokes process
($|1\rangle|n\rangle\rightarrow|2\rangle|n+1\rangle$) is suppressed.
However, in the resonant case, the Stokes process dominates and the
cycle heats the system. The blue (red) vertical downward arrow shows
the cooling (heating) process, which decreases (increases) the
phonon occupancy $n$ of the MR.\label{fig2}}
\end{figure}

The cooling mechanism can be understood as follows. Two localized
states in the DQD ($|1\rangle$ and $|2\rangle$), with energy level
splitting $\hbar\omega_0$, are driven by a microwave field of
frequency $\omega_d$ (Fig.~\ref{fig1}). Similar to the
resolved-sideband cooling of a trapped ion, the DQD resonant
transition frequency is modulated by the oscillation of the MR, and
the absorption spectrum consists of a series of sidebands at
frequencies ($\omega_0-j\omega_m$), where $j=\pm1,\pm2,...$. When
the energy level splitting $\hbar\omega_0$ of the DQD is tuned to
satisfy the lowest sideband condition, i.e.,
$\omega_m=\omega_0-\omega_d$, the excitation of the DQD from the
ground state $|1\rangle$ to the excited state $|2\rangle$ absorbs a
photon of energy $\hbar(\omega_0-\omega_m)$.
The subsequent decay of the DQD via electron tunneling emits a
photon of energy $\hbar\omega_0$. Hence, each scattering process
carries away the MR's vibrational energy by $\hbar\omega_m$, or
reduces the MR's quantum number $n$ by 1. This cooling process is
described by the state transition
$|1\rangle|n\rangle\rightarrow|2\rangle|n-1\rangle\rightarrow|1\rangle
|n-1\rangle$ (see Fig.~\ref{fig2}). A series of cycles of this
process leads to the cooling of the MR. On the other hand, the
reverse process
$|1\rangle|n\rangle\rightarrow|2\rangle|n+1\rangle\rightarrow|1\rangle
|n+1\rangle$, increasing the phonon number, is suppressed because it
is off-resonance. Like in Sisyphus cooling,\cite{GrajcarNP} a cycle
in one direction induces cooling, while the cycle in the other
direction produces heating (see Fig.~\ref{fig2}).


A similar approach of cooling a semiconductor beam by coupling it to
an optical QD was proposed in Ref.~\onlinecite{Wilson04}. Comparing
it to this study, our approach has the following potential
advantages: (i) the cooling system can be fabricated more easily.
The MR here is used as a part of the capacitor, which is easier to
fabricate compared to the proposal in Ref.~\onlinecite{Wilson04}
that embeds an optical QD in a nanoscale beam; (ii) the decay rate
$\Gamma$ of the upper-state of the resonant system (DQD) is just the
rate of the electron tunneling to the electrode, which is tunable by
varying the gate voltage; (iii) with a DQD, it is easy to achieve
the lowest sideband condition, i.e., $\omega_m=\omega_0-\omega_d$,
for cooling the MR by changing the energy level splitting
$\hbar\omega_0$ via the gate voltage.

In typical transport experiments\cite{Gustavsson07,Wiel03} with QDs,
the tunneling rate $\Gamma$ ranges from $10$~kHz to $10$~GHz, while
the fundamental frequency $\omega_m$ of the MR is of the
order\cite{TF08} of $100$~MHz.
The resolved-sideband cooling regime, i.e., $\omega_m\gg\Gamma$, can
be reached by tuning the tunneling rate $\Gamma$. In this regime,
our results show that the steady average phonon occupancy of the MR
can be of the order
of $(\Gamma/\omega_m)^2\ll1$, indicating that the MR can be cooled
to the ground state.

This paper is organized as follows. In Sec.~II, we introduce a model
of the coupled MR-DQD system and derive its effective Hamiltonian.
In Sec.~III, we derive the master equation of the coupled MR-DQD
system and then eliminate the DQD's degrees of freedom to obtain the
master equation for the reduced density matrix of the MR. With this
master equation for the MR, we study, in Sec.~IV, the cooling of the
MR by considering the steady average occupancy of the MR. Moreover,
we analyze the steady phonon occupancy of the MR in the
resolved-sideband cooling regime. Section V summarizes our
conclusions. Furthermore, in Appendices A and B, we show the
derivations of the master equations.
%
%
\section{Mechanical resonator coupled to a double quantum dot}

\subsection{Model}

The circuit diagram of a MR coupled to a DQD is shown in
Fig.~\ref{fig1}(a). The DQD is connected to two electrodes by
tunneling barriers. The bias voltage across the DQD is set such that
the chemical potential of the left electrode $\mu_L$ is higher than
that of the right electrode $\mu_R$ and thus electrons can tunnel
from the left electrode to the right one through the DQD. We assume
that the DQD is in the Coulomb regime, such that at most a single
electron is allowed in the DQD. The corresponding electron states of
the DQD are, respectively, the vacuum state $|0\rangle$, one
electron in the left dot $|1\rangle$, and one electron in the right
dot $|2\rangle$. Here we consider the case where the hopping
strength between the two dots is much smaller than the energy level
splitting of the two dots. To excite the electron from the left dot
to the right one, we apply a microwave field to the DQD [see
Fig.~\ref{fig1}(b)].

As shown in Fig.~\ref{fig1}(a), the capacitor adjacent to the left
dot is formed by a static plate and a single MR with a gate voltage
$V_{g1}$ applied to it. Thus, the displacement of the MR from its
equilibrium position will modulate the capacitance $c_{g1}(x)$. For
typical experimental parameters, the displacement $x$ of the MR is
much less than the equilibrium distance $d$ between the two plates,
i.e., $x\ll{d}$. Hence, the capacitance can be approximately given
by
\begin{equation}
c_{g1}(x)\approx C_{g1}\left(1-\frac{x}{d}\right),
\end{equation}
where $C_{g1}$ is the capacitance at $x=0$. For the harmonically
oscillating MR discussed here, the quantized displacement operator
$x$ can be written as (here $\hbar=1$)
\begin{equation}
x=\sqrt{\frac{1}{2m\omega_m}}(b^{\dagger}+b),
\end{equation}
where $m$ is the effective mass of the MR and $b^+$ ($b$) is the
bosonic creation (annihilation) operator.  The total Hamiltonian of
the whole system reads
\begin{equation}
H_{\rm{total}}\!=\!H_0+H_{\rm{int}}+H_{\rm{T}},
\end{equation}
where
\begin{equation}
H_0\!=\!H_{\rm{leads}}+H_{\rm{DQD}}+H_{\rm{R}},
\end{equation}
is the sum of the isolated bath Hamiltonian $H_{\rm{leads}}$, the
Hamiltonian $H_{\rm{DQD}}$ of the DQD driven by a microwave field,
and the Hamiltonian $H_{\rm{R}}$ of the MR, with
\begin{eqnarray}
H_{\rm{leads}}&=&\sum_{\alpha k}E_{\alpha k}c_{\alpha k}^{\dagger}%
c_{\alpha k},\\
H_{\rm{DQD}}&=&\frac{\omega_0}{2}\sigma_z+\Omega\sigma_x+%
\Omega_0\cos(\omega_dt)\sigma_x,\label{H-DQD}
\\
H_{\rm{R}}&=&\omega_mb^{\dagger}b.
\end{eqnarray}
Here $c_{\alpha k\sigma}^{\dagger}$ ($c_{\alpha k\sigma}$) is the
creation (annihilation) operator of an electron with momentum $k$ in
electrode $\alpha$ $(\alpha=l,r)$.
$\sigma_z=a_2^{\dagger}a_2-a_1^{\dagger}a_1$ and
$\sigma_x=a_2^{\dagger}a_1+~a_1^{\dagger}a_2$ are the Pauli matrices
with $a_1^{\dagger}$ ($a_2^{\dagger}$) being the electron creation
operator in the left (right) dot of the DQD. The second term in
Eq.~(\ref{H-DQD}), $\Omega\sigma_x$, is the hopping tunneling term
between the two dots. The third term in Eq.~(\ref{H-DQD}) describes
the applied microwave field with driving frequency $\omega_d$ and
amplitude $\Omega_0$.

The coupling between the MR and the electron in the left dot is
given by\cite{Amour04}
\begin{eqnarray}
H_{\rm{int}}=-\lambda\;a_1^{\dagger}\,a_1(b^{\dagger}+b),
\end{eqnarray}
with an electromechanical coupling strength $\lambda=\eta\omega_m$.
For a typical electromechanical coupling, $\eta$ $\sim10^{-1}$. The
tunneling coupling between the DQD and the electrodes is
\begin{eqnarray}
H_{\rm{T}}=\sum_{k}(\Omega_{lk}\;a_{1}^{\dagger}\,c_{lk}+\Omega_{rk}\;
a_{2}^{\dagger}\,c_{rk}+\rm{H.c.}),
\end{eqnarray}
where $\Omega_{lk(rk)}$ characterizes the coupling strength between
the QD and the left (right) lead. Hereafter, the subscript ``$l$"
(``$r$") refers to the left (right) electrode.

\subsection{Effective Hamiltonian}

It is difficult to directly analyze the coupled system due to the
different time scales for the dynamics of the DQD, the MR, and the
coupling between them.  In order to solve this problem, we first
eliminate the coupling term between the MR and the DQD through a
canonical transform $U=e^S$ on the whole system, where
\begin{equation}
S=\exp[{-\eta a_1^{\dagger}a_1(b^{\dagger}-b)}].
\end{equation}
With the following relations:
\begin{eqnarray}
Ua_1U^{\dagger}&=&a_1\exp[{\eta(b^{\dagger}-b)}],
\nonumber\\
Ua_2U^{\dagger}&=&a_2,
\nonumber\\
UbU^{\dagger}&=&b+\eta a_1^{\dagger}a_1,
\end{eqnarray}
the transformed Hamiltonian is given by
\begin{eqnarray}
H&=&UH_{\rm{total}}U^{\dagger} \nonumber\\
&=&\sum_{\alpha k}E_{\alpha k}c_{\alpha k}^{\dagger}c_{\alpha
k}+\frac{\omega_0}{2}\sigma_z+\omega_mb^{\dagger}b
\nonumber\\
&&+[\Omega+\Omega_0\cos(\omega_dt)](\sigma_+B^{\dagger}+\rm{H.c.}) \nonumber\\
&&+\sum_{k}(\Omega_{lk}\,a_{1}^{+}c_{lk}B^{\dagger}+\Omega_{rk}\,
a_{2}^{\dagger}\,c_{rk}+\rm{H.c.}),\label{H-Us}
\end{eqnarray}
where we have redefined the energy level splitting $\omega_0$ and
\begin{equation}
B=\exp[{\eta(b^{\dagger}-b)}].\label{B}
\end{equation}
Here we introduce the ladder operators $\sigma_+=a_2^{\dagger}a_1$
and $\sigma_-=\sigma_+^{\dagger}$.

Moreover, to eliminate the time-dependence of the driving term in
Eq.~(\ref{H-Us}), we now employ a unitary transform
\begin{equation}
U_R=\exp\left[{-i\frac{\omega_d}{2}\sigma_zt}\right]
\end{equation}
to change the Hamiltonian to a rotating frame. By neglecting the
fast-oscillating terms within the rotating-wave approximation, the
resulting Hamiltonian is given by
\begin{eqnarray}
H&=&H_{\rm{sys}}+\sum_{\alpha k}E_{\alpha k}c_{\alpha
k}^{\dagger}c_{\alpha k}\nonumber\\
&&+\sum_{k}[\Omega_{lk}\,a_{1}^{+}c_{lk}B^{\dagger}\exp\left({i\frac{\omega_d}{2}t}\right)+\Omega_{rk}\,
a_{2}^{\dagger}c_{rk}+\rm{H.c.}],\nonumber\\
&&~~~~~~~~~
\end{eqnarray}
with
\begin{equation}
H_{\rm{sys}}=-\,\frac{\Delta}{2}\sigma_z+\omega_mb^{\dagger}b%
+\frac{\Omega_0}{2}(\sigma_+B^{\dagger}+\rm{H.c.}).
\end{equation}
Here, $\Delta=\omega_d-\omega_0$ is the driving frequency detuning
the microwave-field  from the transition frequency of the DQD. The
rotating-wave approximation is valid when $\omega_d\gg\Omega$, which
corresponds to a weak hopping tunneling between the two dots.

\section{Master equation for the mechanical resonator}

Within the Born-Markov approximation, by integrating over the
electrode degrees of freedom, we derive a master equation for the
coupled MR-DQD system
\begin{eqnarray}
\frac{d\rho}{dt}=-i[H_{\rm{sys}},
\rho]+\mathcal{L}_{\rm{T}}\rho+\mathcal{L}_{\rm{D}}\rho.\label{ME}
\end{eqnarray}
Here the Liouvillian operator $\mathcal{L}_{\rm{T}}$ presents the
tunneling events through the DQD in the presence of a single MR. In
Eq.~(\ref{ME}) and below, the subscript ``${\rm T}$" (``${\rm D}$")
refers to the tunneling (dissipation). By expanding $B$ in
Eq.~(\ref{B}) up to second order in $\eta$ and assuming that the
energy levels with one-phonon-mediated tunneling are within the bias
window,\cite{Goan04} $\mathcal{L}_{\rm{T}}\rho$ reads
\begin{eqnarray}
\mathcal{L}_{\rm{T}}\rho&=&\Gamma_l(1-\eta^2)\mathcal{D}[a_1^{\dagger}]\rho+\Gamma_r\mathcal{D}[a_2]\rho\nonumber\\
&+&\Gamma_l\eta^2\big(\mathcal{D}[ba_1^{\dagger}]\rho+\mathcal{D}[b^{\dagger}a_1^{\dagger}]\rho\big)\nonumber\\
&+&\Gamma_l\eta^2\big(b^{\dagger}b[a_1,a_1^{\dagger}\rho]+[\rho a_1,
a_1^{\dagger}]b^{\dagger}b\big),
\end{eqnarray}
with the notation $\mathcal{D}$ for any operator $A$:
\begin{equation}
\mathcal{D}[A]\rho=A\rho
A^{\dagger}-\frac{1}{2}[A^{\dagger}A\rho+\rho A^{\dagger}A].
\end{equation}
Here $\Gamma _\alpha=2\pi \rho _\alpha\Omega _{\alpha}^{2}$ is the
rate for electron tunneling to the electrode $\alpha$, while
$\rho_{\alpha}$ denotes the density of states at the electrode
$\alpha$. The Liouvillian operator $\mathcal{L}_{\rm{D}}$ describes
the intrinsic dissipation of the MR induced by its thermal bath and
can be written in a Lindblad form\cite{Scully} as
\begin{eqnarray}
\mathcal{L}_{\rm{D}}\rho&=&\frac{\gamma}{2}[n(\omega_m)+1][2b\rho
b^{\dagger}-(b^{\dagger}b\rho+\rho
b^{\dagger}b)]\nonumber\\
&+&\frac{\gamma}{2}n(\omega_m)[2b^{\dagger}\rho
b-(bb^{\dagger}\rho+\rho bb^{\dagger})],
\end{eqnarray}
where $\gamma=\omega_m/{Q}$ is the intrinsic dissipation rate of the
MR and $n(\omega_m)$ is the average boson number in the thermal
bath.

Next, we focus on the regime in which the driving strength
$\eta\Omega_0$ is low enough so that the time scale related to the
coupling between the MR and the DQD is slow compared to the dynamics
of the DQD and the mechanical oscillation period, just like the
Lamb-Dicke regime considered in laser cooling of an atom or a
trapped ion. We also assume that $[n(\omega_m)+1]\gamma$ is much
smaller than the decay rate of the DQD and the oscillation frequency
of the MR, which is required for appreciable cooling (see Sec.~IV).
In this Lamb-Dicke regime, the DQD can be regarded as a structured
environment and can be adiabatically eliminated.\cite{Cirac92} Since
we are interested in the behavior in the limit $t\rightarrow\infty$,
we can project the system on the subspace with zero eigenvalue of
$\mathcal{L}_0$ ($\mathcal{L}_0$ is the Liouvillian for the
decoupled MR and DQD), according to
\begin{equation}
\mathcal{P}\rho=\rho_d^s\otimes \rm{Tr}_{\it d}\{\rho\}=\rho_{\it
d}^s\otimes \mu,~~~~~~~\mathcal{Q}=1-\mathcal{P},\label{P}
\end{equation}
with $\rho_{d}^s$ denoting the stationary (hence the ``$s$"
superscript) density matrix of the DQD (hence the ``$d$" subscript),
and $\mu$ the density matrix of the MR. Up to second order in
$\eta$, the master equation (\ref{ME}) can be written
as\cite{Cirac92,Wilson07}
\begin{equation}
\frac{d\rho}{dt}=\mathcal{L}(t)\rho=[\mathcal{L}_0(t)+\mathcal{L}_1(t)+\mathcal{L}_2(t)]\rho,\label{ME-DQD-MR}
\end{equation}
where
\begin{eqnarray}
\mathcal{L}_0\rho\!&\!=\!&\!-i[\omega_mb^{\dagger}b,~\rho]%
-i[-\Delta\sigma_z+\frac{\Omega_0}{2}\sigma_x,~\rho] \nonumber\\
&&+\Gamma_l\mathcal{D}[a_1^{\dagger}]\rho+\Gamma_r\mathcal{D}[a_2]\rho,
\\
\mathcal{L}_1\rho\!&\!=\!&\!\mathcal{L}_1^{+}(t)\rho+\mathcal{L}_1^-(t)\rho,
\nonumber\\
\mathcal{L}_1^+\rho\!&\!=\!&\!-i\eta\frac{\Omega_0}{2}[(\sigma_+-\sigma_-)b^{\dagger},~\rho],
\nonumber\\
\mathcal{L}_1^-\rho\!&\!=\!&\!i\eta\frac{\Omega_0}{2}[(\sigma_+-\sigma_-)b,~\rho],
\\
\mathcal{L}_2\rho&=&-\eta^2\Gamma_l\mathcal{D}[a_1^{\dagger}]\rho+\eta^2\Gamma_l%
\big(\mathcal{D}[ba_1^{\dagger}]\rho+\mathcal{D}[b^{\dagger}a_1^{\dagger}]\rho\big)
\nonumber\\
&&+\eta^2\Gamma_l\big(b^{\dagger}b[a_1,a_1^{\dagger}\rho]+[\rho a_1,
a_1^{\dagger}]b^{\dagger}b\big)+\mathcal{L}_{\rm{D}}\rho,\nonumber\\
&&~~~~~~~~~
\end{eqnarray}
are the Liouvillians to zeroth, first, and second order  in $\eta$,
respectively.

Projecting the master equation in Eq.~(\ref{ME-DQD-MR}) on the
$\mathcal{P}$ subspace, one has
\begin{equation}
\frac{d}{dt}\mathcal{P}\rho=[\mathcal{P}\mathcal{L}_2\mathcal{P}%
+\mathcal{P}\mathcal{L}_1(-\mathcal{L}_0)^{-1}\mathcal{L}_1\mathcal{P}\rho].\label{PP}
\end{equation}
This result is obtained to second-order perturbation in $\eta$ (see
Appendix). Since we are interested in the dynamics of the MR, we
will trace Eq.~(\ref{PP}) over the DQD degrees of freedom. The
master equation for the reduced density matrix of the MR is given by
(see Appendix B)
\begin{eqnarray}
\dot{\mu}&=&-i[\omega_m+\delta_m,~b^{\dagger}b]+\frac{1}{2}%
\left\{\gamma[n(\omega_m)+1]+A_-(\omega_m)\right\}
\nonumber\\
&&\times[2b\mu b^{\dagger}-(b^{\dagger}b\mu+\mu b^{\dagger}b)]\nonumber\\
&&+\frac{1}{2}[\gamma n(\omega_m)+A_+(\omega_m)][2b^{\dagger}\mu
b-(bb^{\dagger}\mu+\mu bb^{\dagger})],\nonumber\\
&&~~~~~~~~~~~\label{ME-MR}
\end{eqnarray}
where $\delta_m$ is the driving-induced shift of the mechanical
frequency.

From Eq.~(\ref{ME-MR}), one finds that, besides the effects induced
by the coupling to the thermal bath (terms proportional to
$\gamma$), the cooling and heating induced by the inelastic
scattering processes of the MR can occur and the corresponding rates
$A_{\mp}$ are
\begin{eqnarray}
A_-(\omega)\!&=&\!\frac{\eta^2\Omega_0^2}{2}{\rm{Re}}\bigg\{\frac{1}%
{\Delta^2+(\gamma_0-i\omega)[(\gamma_0-i\omega)+\frac{\Omega_0^2}{2}L(\omega)]}
\nonumber\\
&&\times\big[\frac{\Omega_0}{2}(P(\omega)\langle\sigma_y\rangle_s+iR(\omega)\langle\sigma_x\rangle_s)%
\nonumber\\
&&-(\gamma_0-i\omega)\langle\rho_d^1+\rho_d^2\rangle_s%
+i\Delta\langle\rho_d^1-\rho_d^2\rangle_s\big]\bigg\}+2D,\nonumber\\
A_+(\omega)&=&A_-(-\omega),\label{A-}
\end{eqnarray}
where $\langle\cdot\cdot\cdot\rangle_s$ means the steady state
solution of the corresponding quantity and
\begin{eqnarray}
L(\omega)\!&\!=\!&\!\frac{\Gamma_r+2\Gamma_l-2i\omega}{(\Gamma_l-i\omega)(\Gamma_r-i\omega)},
\nonumber\\
P(\omega)\!&\!=\!&\!\frac{\gamma_0-i\omega}{\Gamma_l-i\omega}%
\left[\frac{2\Gamma_l}{-i\omega}-\frac{(2\Gamma_l-i\omega)}{(\Gamma_r-i\omega)}+1\right],
\nonumber\\
R(\omega)\!&\!=\!&\!\frac{\gamma_0-i\omega}{\Gamma_l-i\omega}%
\left[1+\frac{(2\Gamma_l-i\omega)}{(\Gamma_r-i\omega)}\right],
\nonumber\\
D\!&\!=\!&\!\frac{1}{2}\eta^2\Gamma_l\langle\rho_d^0\rangle_s.
\end{eqnarray}
From Eq.~(\ref{A-}), one can see that the rates $A_{\mp}$ depend on
the tunneling rate $\Gamma_{l(r)}$. For simplicity, below we
consider the symmetric couplings of the DQD to the electrodes, i.e.,
$\Gamma_l=\Gamma_r\equiv\Gamma$.
\begin{figure}
\includegraphics[width=3.0in,
bbllx=24,bblly=10,bburx=167,bbury=245]{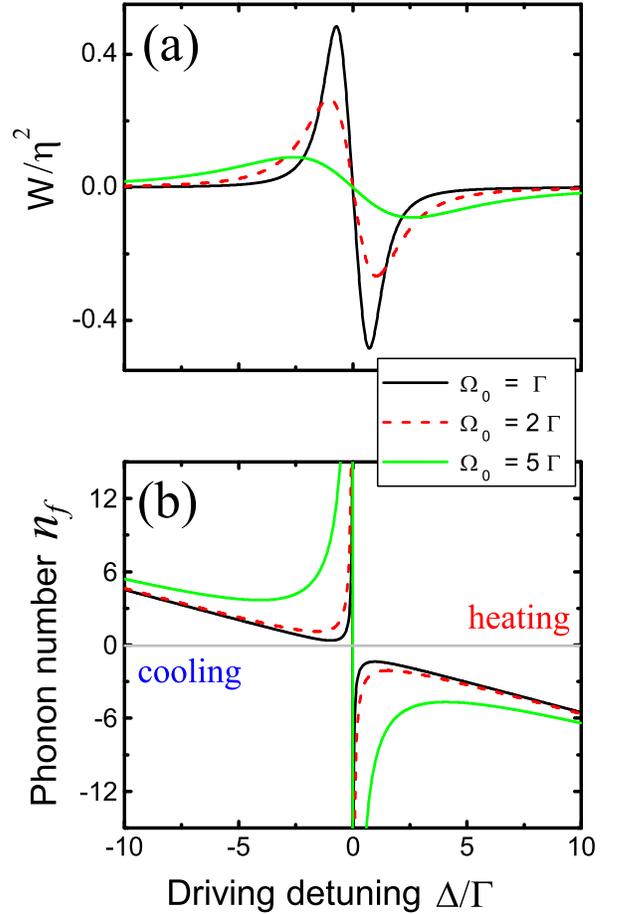} \caption{(Color
online)~(a) The rate $W/\eta^2$ as a function of the driving
detuning $\Delta=\omega_d-\omega_0$, for different driving strengths
$\Omega_0$. For red detuning $\Delta<0$, the rate $W>0$ and the MR
can be cooled. In contrast, the heating process dominates for blue
detuning $\Delta>0$. (b) Microwave-induced steady phonon occupancy
$n_f$ as a function of the normalized driving detuning
$\Delta/\Gamma$. The positive (negative) value of the steady phonon
occupancy on the left (right) side indicates that the scattering
process induced by the microwave produces cooling (heating) of the
MR. The parameters are $\Gamma=1$, $\omega_m=\Gamma$, and
$\Omega_0=\Gamma$~(black), $\Omega_0=2~\Gamma$~(red),
$\Omega_0=5~\Gamma$~(green).\label{fig3}}
\end{figure}
\section{Steady-state average phonon number in the mechanical resonator}

\subsection{Cooling condition}

Below we study the cooling limit regarding the steady-state average
phonon occupancy in the MR. The equation of motion describing the
phonon occupancy distribution can be obtained from the master
equation (\ref{ME-MR}) of the MR, i.e.,
\begin{eqnarray}
\frac{dp_n}{dt}&=&\big\{\gamma[n(\omega_m)+1]+A_-\big\}[(n+1)p_{n+1}-np_n]
\nonumber\\
&&+[\gamma n(\omega_m)+A_+][np_{n-1}-(n+1)p_n],
\end{eqnarray}
where $p_n=\langle n|\mu|n\rangle$. At steady state, $dp_n/dt=0$.
Its solution gives the steady-state average phonon occupancy
\begin{equation}
\langle n\rangle=\frac{\gamma n(\omega_m)+A_+}{\gamma+W},\label{phn}
\end{equation}
where $W=A_--A_+$ is the rate of cooling or heating.

With the expression (\ref{phn}) for the steady-state average phonon
occupancy, we can obtain the cooling condition for the single MR.

To achieve cooling, the condition $W>0$ is required. Otherwise,
cooling the MR is unachievable since the heating processes plays a
dominant role for $W<0$. In Fig.~\ref{fig3}(a), the rate $W$ is
plotted as a function of the driving detuning for different driving
strengths. It can be seen that the sign of the rate $W$ exhibits a
dependence on the detuning $\Delta$. When the driving is red-detuned
($\Delta<0$), the rate $W>0$ and the MR is cooled. The steady-state
average phonon occupancy is plotted in Fig.~\ref{fig3}(b). It
clearly reveals the cooling ($\Delta<0$) and heating ($\Delta>0$)
regions regarding the driving detuning. Below we focus on the
red-detuned region to discuss the cooling of the MR.

At the beginning, if the MR is in its thermal equilibrium state, the
initial average phonon occupancy of the MR is given by
$n(\omega_m)=n_{\rm{th}}$. The numerator of Eq.~(\ref{phn}) reveals
that two parts contribute to the steady-state average phonon
occupancy. The first term, $\gamma n(\omega_m)$, results from the
thermal bath and the steady-state average phonon occupancy is
proportional to the initial thermal occupancy. The second term,
$A_+$, in Eq.~(\ref{phn}) originates from the scattering process
induced by the driving microwave. In order to achieve an appreciable
cooling, i.e., $\langle n\rangle\ll n_{\rm{th}}$, we need
$W\gg\gamma$. In this regime, the steady-state average phonon
occupancy is approximately given by
\begin{equation}
\langle n\rangle\approx n_f=\frac{A_+}{W}.\label{nf}
\end{equation}
\begin{figure}
\includegraphics[width=3.20in,
bbllx=30,bblly=10,bburx=251,bbury=338]{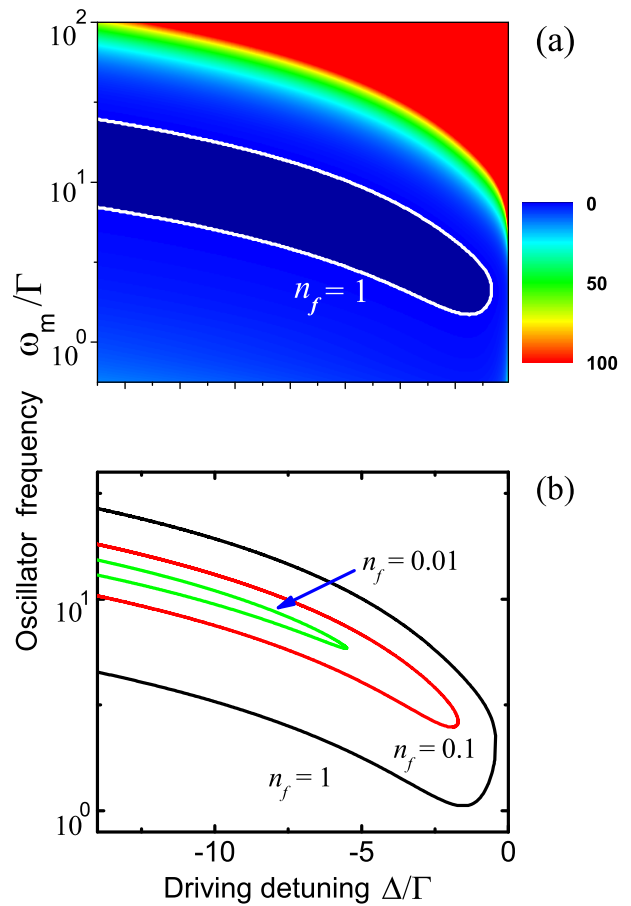} \caption{(Color
online)~(a) Microwave-induced steady-state average phonon occupancy
$n_f$, shown by colors, as a function of both the normalized driving
detuning $\Delta/\Gamma$ and the normalized oscillation frequency
$\omega_m/\Gamma$. (b) Three contour curves of the microwave-induced
steady-state average phonon occupancy for $n_f=1,~0.1$ and $0.01$.
The parameters are $\Gamma=1$, and $\Omega_0=2~\Gamma$.\label{fig4}}
\end{figure}

In Fig.~\ref{fig4}(a), we plot $n_f$ as a function of both the
driving detuning and the oscillation frequency of the MR. The region
for cooling the MR is enclosed by the contour line $n_f=1$. This
region covers a wide area in the $\omega_m$-$\Delta$ plane, implying
that the cooling of the MR is experimentally accessible. Moreover,
the region with steady-state average phonon occupancy much smaller
than unity (e.g., $n_f=0.01\ll1$) is explicitly shown in
Fig.~\ref{fig4}(b). This region corresponds to the cooling of the MR
to the ground state and it can be achieved by both changing the
decay rate of the DQD and detuning the transition frequency of the
DQD from the microwave frequency.

\subsection{Resolved-sideband cooling}

Next, we analytically study the phonon occupancy in the
resolved-sideband cooling region, i.e., $\omega_m\gg\Gamma$. In this
regime, the MR's motional sidebands are well resolved since the
natural linewidth $\Gamma$ of the absorption sidebands for different
mechanical modes are weakly overlapping. This enables highly
targeted cooling with only one mechanical mode.

The previous {\it semiclassical} analyses
\cite{Metzger04,Schliesser06,Brown07,Xue07,Arcizet06} show that the
resonant system (here, the DQD) {\it cannot} respond instantaneously
to the mechanical motion. Hence, this finite response time induces a
phase lag, which produces a force opposing the mechanical motion,
leading to a reduction of the mechanical motion. Here, in {\it
quantum} theory, we will see how the energy is exchanged between the
DQD and the MR during the cooling process (see also
Ref.~\onlinecite{Grajcar08}). We find below that when the DQD is
tuned, via varying the gate voltage, to satisfy the lowest sideband
condition, i.e., $\omega_m=\omega_0-\omega_d$, the MR can be cooled
sufficiently. In this case, the anti-Stokes process is resonantly
enhanced, as discussed in the introduction.

In the resolved-sideband cooling regime, the rates $A_{\mp}$ are
approximately given by
\begin{eqnarray}
&&A_{\mp}\left(\omega_m\right)\;\approx\;\eta
^2\Gamma\left\langle{\rho_d^0} \right\rangle_s +
\frac{1}{2}\frac{\eta ^2\Omega _0^2} {{\left( {\Delta ^2  - \omega_m
^2 + \Omega _0^2 } \right)^2  + \left( {\omega_m
\Gamma } \right)^2 }} \nonumber\\
&&\times
\left[{\frac{\Gamma}{2}\left({\Delta^2+\omega_m^2+\Omega_0^2}
\right) \left\langle {\rho^2_d + \rho^1_d}\right\rangle_s \mp\Delta
\omega_m \Gamma
\langle {\rho^1_d-\rho^2_d}\rangle_s} \right], \hfill \notag\\
&&~~~~~~~~~~\label{A-Sideband}
\end{eqnarray}
and the steady-state average phonon occupancy becomes
\begin{eqnarray}
n_f\!&\!=\!&\!\frac{A_+}{W}=-\frac{1}{2}+\frac{1}{-\Delta\omega_m(4\Delta^2+\Gamma^2)}.\nonumber\\
&&\times\bigg\{(\Delta^2-\omega_m^2+\Omega_0^2)^2+\Gamma^2\omega_m^2%
\nonumber\\
&&+(\Delta^2+\omega_m^2+\Omega_0^2)\left(\Delta^2%
+\frac{\Omega_0^2}{2}+\frac{\Gamma^2}{4}\right)\bigg\}.\label{Phn-Sideband}
\end{eqnarray}
%
%
\begin{figure}
\includegraphics[width=3.20in,
bbllx=20,bblly=10,bburx=185,bbury=248]{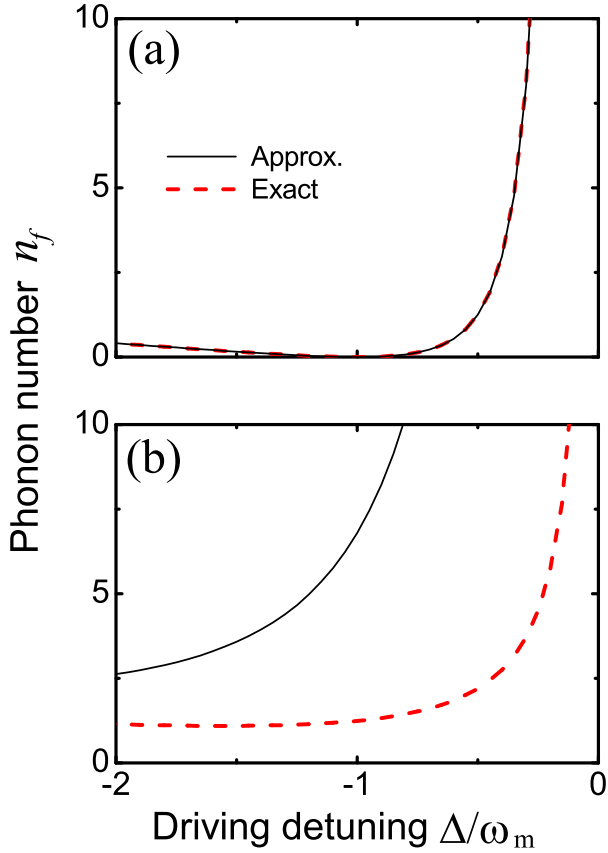} \caption{(Color
online)~Comparison of the steady-state average phonon occupancy
$n_f$ between the exact (\ref{A-}) and approximate
(\ref{A-Sideband}) expressions as a function of the normalized
driving detuning $\Delta/\omega_m$. The parameters are $\Gamma=1$,
$\Omega_0=2~\Gamma$, as well as (a) ${\omega_m}/{\Gamma}=100$, and
(b) ${\omega_m}/{\Gamma}=1$.\label{fig5}}
\end{figure}

As shown in Eq.~(\ref{Phn-Sideband}), the steady-state average
phonon occupancy $n_f$ only depends on the normalized detuning
$\Delta/\Gamma$ for a fixed MR oscillation frequency. With an
optimal detuning value $\Delta=-\omega_m$, the minimum limit of
$n_f$ is found to be
\begin{equation}
n_{\rm{min}}={\rm{min}}\{n_f\}=\frac{7}{8}\bigg(\frac{\Gamma}{\omega_m}\bigg)^2,
\end{equation}
which is much smaller than unity in the resolved-sideband cooling
region, i.e., $\omega_m\gg\Gamma$. This result also shows that when
the DQD is tuned to the lowest sideband $\Delta=-\omega_m$, the
anti-Stokes process
($|1\rangle|n\rangle\rightarrow|2\rangle|n-1\rangle$) is resonantly
enhanced. Therefore, an appreciable cooling is achieved. The result
obtained here is also consistent with previous theoretical
predictions for the resolved-sideband cooling
limit,\cite{Wineland87,Wilson07} which has been verified by
experiments.\cite{Diedrich89,ASchliesser08} However, an important
advantage in the present set-up is that both the decay rate and the
energy splitting of the DQD can be tuned by varying the gate voltage
to reach the resolved-sideband cooling regime. In this regime, the
steady-state average phonon occupancy is much smaller than unity.
Thus, the MR can be cooled to its ground state.

Moreover, in Fig.~\ref{fig5} we plot the steady-state average phonon
occupancy in the appreciable cooling regime using the exact
[Eq.~(\ref{A-})] and the approximate expressions
[Eq.~(\ref{A-Sideband})] for the rates $A_{\mp}$, respectively. In
the resolved-sideband regime, the results agree well with each other
in these two cases [Fig.~\ref{fig5}(a)]. In Fig.~\ref{fig5}(b), when
the coupled MR-DQD system deviates from the resolved-sideband
regime, however, the exact and the approximate expressions can
differ significantly from each other. This regime implies the
breakdown of the approximation used to obtain
Eq.~(\ref{A-Sideband}). This indicates that one {\it needs} to use
the {\it exact} expression (\ref{A-}) {\it to describe the MR
cooling in the non-resolved-sideband regime}. Indeed, as shown in
Fig.~\ref{fig4}, though the condition $\omega_m\gg\Gamma$ is not
fulfilled in the non-resolved-sideband regime, the ground-state
cooling of the MR is still achievable.

\subsection{Estimates}

Finally, let us estimate the steady-state average phonon occupancy
of the MR in the resolved-sideband regime using typical experimental
parameters.\cite{ASchliesser08,Gustavsson07,Wiel03} Here we use
$\omega_m=2\pi\times100$~MHz, $\Gamma=2\pi\times10$~MHz,
$\Omega_0=2\pi\times20$~MHz, and $\eta=0.2$. When the DQD is tuned
to the lowest sideband $\Delta=-\omega_m$, the microwave-induced
steady-state average phonon occupancy is given by
\begin{equation}
n_f=\frac{7}{8}\bigg(\frac{\Gamma}{\omega_m}\bigg)^2=0.00875,
\end{equation}
and the cooling rate is
\begin{equation}
W=A_--A_+\approx1.4~{\rm MHz}.
\end{equation}
Considering a MR with a quality factor $Q=10^5$, the intrinsic
dissipation rate of the MR is $\gamma=1$~kHz. Hence, an appreciable
cooling effect, i.e., $W\gg\gamma$, can be produced. For a MR
precooled by a dilution refrigerator to a temperature $T_0$ of,
e.g., $100$~mK, we have $n_{\rm{th}}\approx21$, and thus the
steady-state average phonon occupancy part that comes from thermal
fluctuations is
\begin{equation}
\frac{\gamma\;n_{\rm{th}}}{W}\approx0.015.
\end{equation}
It follows from Eq.~(\ref{phn}) that the steady-state average phonon
occupancy of the MR is given by
\begin{equation} \langle
n\rangle\approx0.024\ll1,\label{Estimate-n}
\end{equation}
which corresponds to an effective temperature $T_{\rm
eff}\approx1.3$~mK. This means that the cooling of the MR to the
ground state is achievable using the proposed set-up with a MR
coupled to a DQD.

As derived in Ref.~\onlinecite{Grajcar08}, starting from an initial
temperature $T_0$, the final temperature of the cooled MR is bound
by
\begin{equation}
T_f=\frac{\omega_m}{\omega_0}\;T_0,\label{T-Limit}
\end{equation}
i.e., this expression provides the lower limit of the temperature
that can be achieved via the sideband cooling. Experimentally, the
energy level difference $\Delta\varepsilon$ between the ground state
and the first excited state of a single quantum dot can be
$\sim250~\mu$eV (see, e.g., Ref.~\onlinecite{Onac06}), which
corresponds to a frequency $\nu\sim60~$GHz. This level difference
can be even larger by decreasing the size of the dot. In our set-up,
$\Delta\varepsilon$ should be larger than the energy level
difference $\hbar\omega_0$ between the ground state of the left dot
and that of the right dot, so as to prevent the electron in the
right dot from tunneling to the first excited state of the left dot.
Here, for example, we can choose $\omega_0=2\pi\times40$~GHz. Other
parameters used for calculating $\langle n\rangle$ in
Eq.~(\ref{Estimate-n}) are $\omega_m=2\pi\times100$~MHz, and
$T_0=100$~mK. From Eq.~(\ref{T-Limit}), we have $T_f=0.25$~mK.
Obviously, the temperature limit is lower than the achieved
temperature $T_{\rm eff}\approx1.3$~mK in Eq.~({\ref{Estimate-n}}).
This implies that the MR can be further cooled using our sideband
cooling proposal.

\section{Conclusion}

We have studied the cooling of a MR by electrostatically coupling it
to a semiconductor DQD. Here the DQD works as a two-level system and
the decay rate of the DQD corresponds to the rate of the electron in
the DQD tunneling to the electrode. This tunneling rate and the
energy level splitting of the DQD can be tuned by varying the gate
voltage. We show that when the two-level system is driven by a
microwave field in red-detuning, the MR can be cooled, in analogy to
the laser sideband-cooling of atoms or trapped ions in atomic
physics. Also, we obtain analytical results for the
resolved-sideband cooling of the MR. Moreover, our results show that
the ground-state cooling of the MR can be achieved both by detuning
the transition frequency of the DQD from the microwave frequency and
by changing the decay rate of the DQD. Importantly, this frequency
detuning and the decay rate of the DQD are tunable by varying the
gate voltages of the DQD. Thus, the coupled MR-DQD system provides
an experimentally implementable set-up for ground-state cooling of
MRs.

\begin{acknowledgments}
We thank S. Ashhab for useful discussions. F.N. was supported in
part by the National Security Agency (NSA), the Laboratory for
Physical Sciences (LPS), the Army Research Office (ARO), and the
National Science Foundation (NSF) Grant No. EIA-0130383. S.H.O. and
J.Q.Y. were supported by the National Fundamental Research Program
of China grant No. 2006CB921205 and the National Natural Science
Foundation of China grant Nos. 10534060 and 10625416.
\end{acknowledgments}

\appendix
\section{Master equation for the density matrix projected on a subspace of the Liouvillian $\mathcal{L}_0$}

In this appendix, we derive Eq.~(\ref{PP}) by projecting the master
equation (\ref{ME}) on the subspace, with zero eigenvalue
$\lambda_0=0$, of the Liouvillian $\mathcal{L}_0$ for the decoupled
resonator and DQD system. The projection is defined as
\begin{equation}
\mathcal{P}\rho=\rho_d^{s}\otimes {\rm Tr}_{\it
d}\{\rho\}=\rho_d^{s}\otimes
\mu,~~~~~~~\mathcal{Q}=1-\mathcal{P},\label{P}
\end{equation}
where $\mathcal{P}$ is the projection operator. The definition of
$\mathcal{P}$ implies that
$\mathcal{L}_0\mathcal{P}=\mathcal{P}\mathcal{L}_0=0$, which leads
to
\begin{eqnarray}
\mathcal{P}\mathcal{L}_0\mathcal{P}=\mathcal{Q}\mathcal{L}_0\mathcal{P}=\mathcal{P}\mathcal{L}_0\mathcal{Q}=0,
~~\mathcal{Q}\mathcal{L}_0\mathcal{Q}=\mathcal{L}_0.\label{Relat1}
\end{eqnarray}
With these relations, the projection of the master equation
(\ref{ME}) gives
\begin{eqnarray}
\mathcal{P}\dot{\rho}&=&\mathcal{P}\mathcal{L}_2\mathcal{P}\rho%
+\mathcal{P}(\mathcal{L}_1+\mathcal{L}_2)\mathcal{Q}\rho,
\\
\mathcal{Q}\dot{\rho}&=&[\mathcal{L}_0+\mathcal{Q}(\mathcal{L}_1+\mathcal{L}_2)]\mathcal{Q}\rho%
+\mathcal{Q}(\mathcal{L}_1+\mathcal{L}_2)\mathcal{P}\rho.\label{PME}
\nonumber\\
&&~~~
\end{eqnarray}
Here we have used the relation
$\mathcal{P}\mathcal{L}_1\mathcal{P}=0$, due to the fact that
tracing over the interaction between the DQD and the MR equals zero.
Next, we define
\begin{eqnarray}
v(t)\equiv\mathcal{P}\rho(t),~w(t)\equiv\mathcal{Q}\rho(t).
\end{eqnarray}
Applying the Laplace transform
\begin{equation}
\tilde{b}(s)=\int\limits_0^\infty b(t)e^{-st}dt
\end{equation}
on Eq.~(\ref{PME}), one has
\begin{eqnarray}
s\tilde{v}(s)-v(0)\!&\!=\!&\!\mathcal{P}\mathcal{L}_2\tilde{v}(s)+\mathcal{P}(\mathcal{L}_1+\mathcal{L}_2)\tilde{w}(s),
\nonumber\\
s\tilde{w}(s)-w(0)\!&\!=\!&\![\mathcal{L}_0+\mathcal{Q}(\mathcal{L}_1+\mathcal{L}_2)]\tilde{w}(s)%
\nonumber\\
&&+\mathcal{Q}(\mathcal{L}_1+\mathcal{L}_2)\tilde{v}(s). \label{LP}
\end{eqnarray}

Then we introduce a small parameter $\zeta$ to characterize the
order of the Liouvillians,\cite{Zoller} i.e.,
\begin{equation}
\mathcal{L}_0(t)\rightarrow\mathcal{L}_0(t),
~\mathcal{L}_1\rightarrow\zeta\mathcal{L}_1(t),
~\mathcal{L}_2\rightarrow\zeta^2\mathcal{L}_2(t).\label{order}
\end{equation}
Substituting Eq.~(\ref{order}) into Eq.~(\ref{LP}) and including
terms up to second-order, one has
\begin{eqnarray}
&&s\tilde{v}(s)-[v(0)+\mathcal{P}(\zeta\mathcal{L}_1+\zeta^2\mathcal{L}_2)(s-\mathcal{L}_0)^{-1}w(0)]
\nonumber\\
&&=\zeta^2\mathcal{P}\mathcal{L}_2\tilde{v}(s)+\zeta^2\mathcal{P}\mathcal{L}_1(s-\mathcal{L}_0)^{-1}\mathcal{Q}\mathcal{L}_1
\tilde{v}(s).
\end{eqnarray}
Finally, neglecting the correction due to the initial condition and
performing the inverse Laplace transform, one obtains
\begin{eqnarray}
\mathcal{P}\dot{\rho}=\mathcal{P}\mathcal{L}_2\mathcal{P}\rho
+\mathcal{P}\mathcal{L}_1\mathcal{Q}(-\mathcal{L}_0)^{-1}\mathcal{Q}\mathcal{L}_1\mathcal{P}\rho,\label{APP}
\end{eqnarray}
which is just Eq.~(\ref{PP}) in Sec.~III.

\section{Master equation for the reduced density matrix of the mechanical resonator}

Below we derive the master equation for the reduced density matrix
$\mu$ of the MR. Following the procedures in
Ref.~\onlinecite{Cirac92}, we trace over the DQD degrees of freedom
in Eq.~(\ref{APP}), the first term on the r.h.s.~of Eq.~(\ref{APP})
gives
\begin{eqnarray}
&&\rm{Tr}_{\it d}({\mathcal{P}\mathcal{L}_2\mathcal{P}})
=-i\frac{\eta^2\Omega_0}{2}\langle{\sigma _{\it x}}\rangle_{\it
s}[{\it b^{\dagger}b},\mu]
\nonumber\\
&&+\frac{1}{2}\big\{\gamma[n{(\omega_m)}+1]+\eta^2\Gamma_l
\langle\rho_{d}^0\rangle_{s}\big\}
\nonumber\\
&&\times[2b\mu b^{\dagger}-(b^{\dagger}b\mu+\mu b^{\dagger}b)]
\nonumber\\
&&+\frac{1}{2}[\gamma
n(\omega_m)+\eta^2\Gamma_l\langle\rho_{d}^0\rangle_{s}][2b^{\dagger}\mu
b -(bb^{\dagger}\mu+ \mu bb^{\dagger})]. \nonumber\\
&&~~~~~~~~~~~~~\label{ME-1}
\end{eqnarray}
The second term on the r.h.s.~of Eq.~(\ref{APP}) gives
\begin{eqnarray}
&&\rm{Tr}_{\it
d}\{\mathcal{P}\mathcal{L}_1(-\mathcal{L}_0)^{-1}\mathcal{L}_1\mathcal{P}\rho\}
\!=\!\int\limits_0^{+\infty} {\it dt}\;
Tr_d\{\mathcal{P}\mathcal{L}_1e^{-\mathcal{L}_0t}\mathcal{L}_1\mathcal{P}\rho\}
\nonumber\\
&&\!=\!-i~{\rm{Im}}[S(\omega_m)+S(-\omega_m)][b^{\dagger}b,\mu] \nonumber\\
&&+~{\rm{Re}}[S(\omega_m)][2b\mu b^{\dagger}-(b^{\dagger}b\mu+\mu b^{\dagger}b)]\nonumber\\
&&+~{\rm{Re}}[S(-\omega_m)][2b^{\dagger}\mu b-(bb^{\dagger}\mu+\mu
bb^{\dagger})],\label{ME-2}
\end{eqnarray}
where
\begin{eqnarray}
S(\omega)=\eta^2\frac{\Omega_0^2}{4}\int\limits_0^{+\infty}dt%
\;e^{i\omega t}\langle\sigma_y(t)\sigma_y(0)\rangle.\label{CF}
\end{eqnarray}
Substituting Eqs.~(\ref{ME-1}), (\ref{ME-2}), and (\ref{P}) into
(\ref{APP}), we obtain the master equation for the MR
\begin{eqnarray}
\dot{\mu}\!&=&\!-i[\omega_m+\delta_m,~b^{\dagger}b]+\frac{1}{2}%
\left\{\gamma[n(\omega_m)+1]+A_-(\omega_m)\right\}
\nonumber\\
&&\times[2b\mu b^{\dagger}-(b^{\dagger}b\mu+\mu b^{\dagger}b)]\nonumber\\
&&+\frac{1}{2}[\gamma n(\omega_m)+A_+(\omega_m)][2b^{\dagger}\mu
b-(bb^{\dagger}\mu+\mu bb^{\dagger})].\nonumber\\
&&~~~~~~~~~~~\label{AME-MR}
\end{eqnarray}
This is just Eq.~(\ref{ME-MR}) in Sec.~III.

In Eq.~(\ref{AME-MR}), we have introduced
\begin{equation}
\delta_m=\eta^2\frac{\Omega_0}{2}\langle\sigma_x\rangle_s%
+{\rm{Im}}[S(\omega_m)+S(-\omega_m)]
\end{equation}
and the rates $A_{\mp}(\omega_m)$
\begin{eqnarray}
A_{\mp}(\omega_m)=2~{\rm{Re}}[S(\pm\omega_m)+D],\label{AA-}
\end{eqnarray}
with
\begin{equation}
D=\frac{1}{2}\eta^2\Gamma_l\langle\rho_d^0\rangle_s,
\end{equation}
where $\langle\rho_d^0\rangle_s$ is the probability of an empty DQD
at the steady state.

To determine the rates $A_{\mp}$, we need to calculate the
correlation function $S(\omega)$ of Eq.~(\ref{CF}) using the
equation of motion for the DQD:
\begin{equation}
\dot{\rho}_d=-i[\Delta\sigma_z+\frac{\Omega_0}{2}\sigma_x,~\rho_d]%
+\Gamma_l\mathcal{D}[a_1^{\dagger}]\rho_d+\Gamma_r\mathcal{D}[a_2]\rho_d.\label{ME-DQD}
\end{equation}
From Eq.~(\ref{ME-DQD}), one can obtain the following equations of
motion:
\begin{eqnarray}
\dot{\rho_d^1}&=&\Gamma_l-\Gamma_l\rho_d^1%
-\Gamma_l\rho_d^2-\frac{\Omega_0}{2}\langle\sigma_y\rangle,
\nonumber\\
\dot{\rho_d^2}&=&-\Gamma_r\rho_d^2+\frac{\Omega_0}{2}\langle\sigma_y\rangle,
\nonumber\\
\dot{\langle\sigma_x\rangle}&=&\Delta\langle\sigma_y\rangle%
-\gamma_0\langle\sigma_x\rangle,
\nonumber\\
\dot{\langle\sigma_y\rangle}&=&-\Delta\langle\sigma_x\rangle-\Omega_0(\rho_d^2-\rho_d^1)%
-\gamma_0\langle\sigma_y\rangle.\label{EOM}
\end{eqnarray}
where $\gamma_0=\Gamma_r/2$. With the normalization condition
$\rho_d^0+\rho_d^1+\rho_d^2=1$, the steady-state solution of
Eq.~(\ref{ME-DQD}) is given by
\begin{eqnarray}
&&\langle\rho_d^0\rangle_s=\frac{\Omega_0^2\Gamma_r}{M},~~~~~~~
\langle\rho_d^1\rangle_s=\frac{4\Gamma_l\Delta^2+\Gamma_l\Omega_0^2+\Gamma_l\Gamma_r^2}{M},
\nonumber\\
&&\langle\rho_d^2\rangle_s=\frac{\Gamma_l\Omega_0^2}{M},~~~~~~~~
\langle\sigma_x\rangle_s=\frac{4\Gamma_l\Delta\Omega_0}{M},\nonumber\\
&&\langle\sigma_y\rangle_s=\frac{2\Gamma_l\Gamma_r\Omega_0}{M},
\end{eqnarray}
where
\begin{equation}
M=4\Gamma_l\Delta^2+\Omega_0^2(2\Gamma_l+\Gamma_r)+\Gamma_l\Gamma_r^2.
\end{equation}
From Eq.~(\ref{EOM}), one can obtain
\begin{eqnarray}
\langle\tilde{\sigma}_y(s)\rangle\!&\!=\!&\!\left\{\Delta^2+(s+\gamma)^2+\frac{\Omega_0^2}{2}(s+\gamma)%
\frac{2s+2\Gamma_l+\Gamma_r}{(s+\Gamma_l)(s+\Gamma_r)}\right\}^{-1}\nonumber\\
&&\times\bigg\{\Omega_0\frac{\Gamma_l(s+\gamma)}{s(s+\Gamma_l)}%
+\Omega_0\frac{(s+\gamma)}{(s+\Gamma_l)}\rho_d^1(0)%
\nonumber\\
&&-\Omega_0\frac{(s+\gamma)(s+2\Gamma_l)}{(s+\Gamma_l)(s+\Gamma_r)}\rho_d^2(0)%
\nonumber\\
&&+(s+\gamma)\langle\sigma_y\rangle(0)-\Delta\langle\sigma_x\rangle(0)\bigg\},
\end{eqnarray}
which is the Laplace transform of $\langle{\sigma}_y(t)\rangle$.
Then, using the quantum regression theorem,\cite{Scully} and
performing the inverse Laplace transform, one can calculate the
correlation function in Eq.~(\ref{CF}). Subsequently, with
Eq.~(\ref{AA-}), the rates $A_{\mp}(\omega_m)$ are obtained as in
Eq.~(\ref{A-}).


\end{document}